\documentclass[aps,twocolumn,showpacs,preprintnumbers,amsmath,amssymb]{revtex4}

\usepackage{graphicx}
\usepackage{dcolumn}
\usepackage{bm}

\renewcommand{\vec}[1]{\bm{#1}}
\newcommand{\ket}[1]{\left |\mbox{$#1$}\right\rangle}

\newcommand{\ave}[1]{\left\langle\mbox{$#1$}\right \rangle}
\newcommand{\tr}{\mathrm{tr}}
\newcommand{\etal}{{\it et al.}}

\begin{document}

\title{Decoherence of localized spins interacting via RKKY interaction}
\author{Yoshiaki Rikitake and Hiroshi Imamura}
\affiliation{
CREST and Graduate School of Engineering, Tohoku University, Sendai
980-8579, Japan
}


\begin{abstract}
 We theoretically study decoherence of
 two localized spins interacting via the RKKY interaction
 in one-, two-, and three-dimensional electron gas.
 We derive the kinetic equation for the reduced density matrix of 
 the localized spins and show that energy relaxation caused by
 singlet-triplet transition is suppressed when the RKKY interaction is
 ferromagnetic.  We also estimate the decoherence time of the system
 consisting of two
 quantum dots embedded in a two dimensional electron gas.

\end{abstract}

\maketitle

Quantum computation and quantum information are emerging research
fields of physics, technology and information sciences\cite{nielsen}.
The elementary units in most quantum computation and quantum
information schemes are a quantum bit and a quantum gate.
Because of its scalability and 
relatively long coherence time\cite{Kroutvar04},
solid state device with localized spins is considered as
one of the promising candidates for these quantum devices\cite{awshalom}.
Kane\cite{kane98} proposed the system of
nuclear spins of phosphorus donors
in a silicon heterostructure with
direct exchange interaction between electrons
localized at the donors.
Mozyrsky {\it et al.}\cite{mozyrsky01a,mozyrsky01b} proposed the qubits of
nuclear spins in quantum-Hall system in which
the gate operation is realized by
the indirect exchange interaction via virtually excited spin waves.
Recently Craig \etal \cite{craig04,glazman04} observed the RKKY
\cite{kittel} coupling of semiconductor quantum dots, which proved
that the RKKY interaction can be used as a quantum gate consisting of 
localized spins in semiconductor quantum dots.  Several theoretical
papers studying the RKKY interaction in such semiconductor
nanostructures have been
published\cite{piermarocchi02,tamura04,imamura04,utsumi04,sun04,vavilov05}. 

One of the major obstacle to realizing quantum computation is 
decoherence\cite{zurek91,weiss}.
Since the RKKY interaction is mediated by the electron gas,
the particle-hole excitations act as an environment (Fermion
bath)\cite{chang85,chen87a,chen87b,chen88}.
It is therefore important to clarify the
effects of the Fermion bath on the dynamics of the qubits consisting
of the localized spins.

In this paper, we study the dynamics of the localized spins
interacting via the RKKY interaction by using the kinetic equation of
the reduced density matrix. 
We find the term intrinsic to the RKKY interaction
appears in the kinetic equation, which is an oscillating function of
the distance between the localized
spins with the same period as the RKKY interaction. 
We show that the energy relaxation due to the singlet-triplet
transition is strongly suppressed
by this term.
We also discuss the physical realization of a quantum gate using the
RKKY interaction and estimate the decoherence time of the system
consisting of two quantum dots embedded in a two dimensional electron
gas(2DEG).

We consider the system consisting of two localized spins
embedded in a one-, two-, or three-dimensional electron gas.
The Hamiltonian,
$
 H = H_{S} + H_{\mathrm{c}} + H_{\mathrm{int}}
$,
comprises the localized spin part, 
$H_{\mathrm{S}}$, the conduction electron part, 
$H_{\mathrm{c}}$, and the $s$-$d$ interaction, 
$H_{\mathrm{int}}$.  
We assume that there is no external magnetic field 
and we set $H_{\mathrm{S}} = 0$.
In the second quantization representation, 
$H_{\mathrm{c}}$ and $H_{\mathrm{int}}$ are 
expressed as
\begin{align}
 H_{\mathrm{c}}
  &=
 \sum_{\vec{k}\sigma}
 \varepsilon_{k}c^{\dag}_{\vec{k}\sigma}c_{\vec{k}\sigma},\\
  H_{\mathrm{int}}
  &=\frac{J}{2V}\sum_{p=1,2}\sum_{\vec{k},\vec{k}^{\prime}}
  e^{i(\vec{k}-\vec{k}^{\prime})\cdot \vec{R}_{p}}
  \Bigl\{
  S^{-}_{p}c^{\dag}_{\vec{k}^{\prime}\uparrow}c_{\vec{k}\downarrow}
\\\nonumber&
  +S^{+}_{p}c^{\dag}_{\vec{k}^{\prime}\downarrow}c_{\vec{k}\uparrow}
  +S^{z}_{p}\left(
  c^{\dag}_{\vec{k}^{\prime}\uparrow}c_{\vec{k}\uparrow}-
  c^{\dag}_{\vec{k}^{\prime}\downarrow}c_{\vec{k}\downarrow}
  \right)\Bigr\},
 \end{align}
where $c_{\vec{k}\sigma}^{\dag}$ and $c_{\vec{k}\sigma}$ are 
creation and annihilation operators 
of an electron with wavenumber vector $\vec{k}$ and spin $\sigma$.  
Here $\varepsilon_{k}$ is an energy of conduction electrons,
$J$ is a coupling constant of $s$-$d$ interaction,
and $\vec{R}_p$ ($p=1,2$) represents the position 
of the localized spin $\vec{S}_p$.
We assume that the coupling constant $J$ is so small 
that we can treat $H_{\mathrm{int}}$ 
as a perturbation on $H_{\mathrm{c}}$.


\begin{table*}[]
 \begin{equation*}
  \begin{array}{c|c|c|c|c}
   \hline
    d & \alpha_d & \beta_d & F_d(x) & G_d(x) \\
   \hline\hline
    1 \rule[-12pt]{0pt}{30pt}
    & \dfrac{m^2J^2}{2\pi\hbar^4k_F^2}
    & 2 \mathrm{si}(\pi)=5.62\times 10^{-1}
    & \dfrac{\mathrm{si}(x)}{\mathrm{si}(\pi)}
    & \dfrac{1+\cos x}{2}
    \\ 
   \hline
    2 \rule[-12pt]{0pt}{30pt}
    & \dfrac{m^2J^2}{32\pi^2\hbar^4}
    & \dfrac{8}{\pi}J_1(z_0)N_1(z_0)=1.10\times 10^{-1}
    & \dfrac{J_0(x/2)N_0(x/2)+J_1(x/2)N_1(x/2)}{J_1(z_0)N_1(z_0)}
    & J_0^2(x/2)
    \\ 
   \hline 
    3 \rule[-12pt]{0pt}{30pt}
    & \dfrac{m^2J^2k_F^2}{16\pi^3\hbar^4}
    & \dfrac{1}{\pi^3}=3.23\times 10^{-2}
    & (2\pi)^3\dfrac{x\cos x-\sin x}{x^4}
    & 2\dfrac{1-\cos x}{x^2}
    \\ 
   \hline 
  \end{array}
 \end{equation*}
 \caption{The coefficients $\alpha_d$ and $\beta_d$, and
 the range functions $F_d(x)$ and $G_d(x)$ 
 for $d$-dimensional electron gas $(d=1,2,3)$. 
 $\mathrm{si}(x)\equiv -\int_x^{\infty}dt\sin t/t$
 is the sine integral function.
 $J_n$ is the Bessel function and
 $N_n$ is the Neumann function of order $n$.
 $z_0=2.40$ is the first zero-point of $J_0$.}
 \label{tb:kappa-gamma}
\end{table*}

The dynamics of the two localized spins 
is described by the reduced density matrix
$\rho(t)=\tr_{\mathrm{c}}\{\rho_{\mathrm{tot}}(t)\}$, 
where $\rho_{\mathrm{tot}}$ is the density matrix for the total system
and $\tr_{\mathrm{c}}$ means trace over the degrees of
freedom of the conduction electrons.
The reduced density matrix $\rho(t)$ obeys 
the following kinetic equation
\cite{schoeller94,Keil01,takagahara,makhlin04}:
\begin{equation}
 \frac{d}{dt}\rho(t)
 =
 -\dfrac{i}{\hbar}\left[H_{S},\rho(t)\right]
 +\int_0^{t}dt^{\prime}\Sigma(t-t^{\prime})\rho(t^{\prime}).
 \label{eq:kineticEq}
\end{equation}
The self-energy  $\Sigma(t-t^{\prime})$ is 
a super-operator acting on $\rho(t)$.
We assume that at the initial time $t=0$
the system of the localized spins 
and the system of the conduction electrons are decoupled,
and the conduction electrons are in thermal equilibrium:
$\rho_{\mathrm{tot}}(0)=\rho(0)\otimes\rho^{eq}_{\mathrm{c}}$, 
where $\rho^{eq}_{\mathrm{c}}=e^{-(H_{\mathrm{c}}-\mu N)/k_BT}/
\tr_{\mathrm{c}}e^{-(H_{\mathrm{c}}-\mu N)/k_BT}$.

We carry out a second-order
perturbative calculation for the self-energy $\Sigma(t-t^{\prime})$ in
Eq.\eqref{eq:kineticEq}.
We assume that the reduced density matrix $\rho(t)$
varies slowly compared to the lifetime of
particle-hole excitations,
and we make the Markov approximation
in Eq.\eqref{eq:kineticEq}.
A straight forward calculation gives 
\begin{equation}
  \dfrac{d}{dt}\rho(t)
  =
  \left(
  \mathcal{H}_{\mathrm{RKKY}}
  +\sum_{p=1,2}\mathcal{D}_p
  +\mathcal{D}_{\mathrm{ex}}  \right)\rho(t),
\label{eq:MasterEq}
\end{equation}
where $\mathcal{H}_{\mathrm{RKKY}}$, 
$\mathcal{D}_p$ $(p=1,2)$, and
$\mathcal{D}_{\mathrm{ex}}$
are the super-operators defined as follows.

\begin{figure}[]
 \begin{center}
  \includegraphics[width=0.85\linewidth,clip]{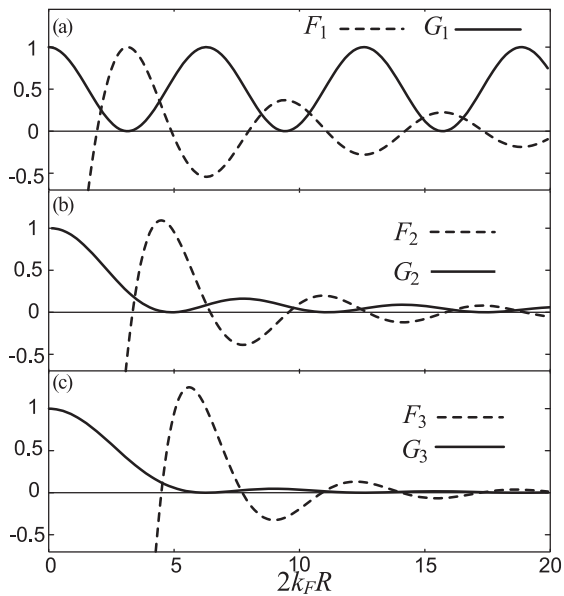}
 \end{center}
 \caption{The range functions for one-dimensional system $F_1(2k_FR)$
 and $G_1(2k_FR)$ are plotted as a function of $2k_{F}R$ in Panel (a).
 Panels (b) and (c) are the same plot for two- and
 three-dimensional systems, respectively.
 }
 \label{fig:rangefunc}
\end{figure}

The first term of r.h.s. in Eq.\eqref{eq:MasterEq} is defined as
\begin{equation}
\mathcal{H}_{\mathrm{RKKY}}\rho(t) =
 -i\dfrac{J_{\mathrm{RKKY}}}{\hbar}
 \left[\vec{S}_1\cdot\vec{S}_2,\rho(t)\right],
 \label{eq:super-op-HRKKY}
\end{equation}
which represents the coherent time evolution of the two localized spins
interacting via the RKKY interaction\cite{kittel,fischer75,yafet87,litvinov98}.
The effective coupling constant $J_{\mathrm{RKKY}}$,
for $d$-dimensional system ($d=1,2,3$) is given by
$J_{\mathrm{RKKY}}=\alpha_d \beta_d E_F  F_d(2 k_FR)$,
where $R=|\vec{R}_1-\vec{R}_2|$ is the distance 
between two localized spins, $E_F$ is the Fermi energy, and 
$k_F$ is the Fermi wavenumber.
Here $\alpha_d$, $\beta_d$, and the range function $F_d(x)$ are given in
Table \ref{tb:kappa-gamma}.  The range functions $F_d(2k_FR)$ for
$d=1,2$, and $3$ are plotted by dotted lines in Figs.\ref{fig:rangefunc}(a),
(b), and (c), respectively.
The RKKY interaction is produced by virtually excited quantum states
within a wide energy window from the bottom of energy band to
the Fermi surface.  Therefore, the strength of the RKKY interaction,
$J_{\mathrm{RKKY}}$, is proportional to the Fermi energy
$E_F$ and is oscillating function of $R$ with the period of half of the
Fermi wavelength, $\pi/k_F$.

The second term of r.h.s. in Eq.\eqref{eq:MasterEq} is defined as
\begin{equation}
 \begin{aligned}
  \sum_{p=1,2}
  \mathcal{D}_p\rho(t)
  &=
  -\dfrac{\gamma}{\hbar}
  \sum_{p=1,2}
  \Bigl(\dfrac{3}{2}\rho(t) 
  -\bigl(
  S_p^+\rho(t)S_p^-
  \\& +
  S_p^-\rho(t)S_p^+
  +
  2 S_p^z\rho(t)S_p^z
  \bigr)
 \Bigr),
 \label{eq:super-op-Dp}
 \end{aligned}
\end{equation}
where the coefficient $\gamma$ is expressed as $\gamma=\alpha_d k_B T$.
This term describes the usual decoherence of a localized spin $\vec{S}_{p}$
interacting with the Fermion bath, which is known as the Korringa
relaxation\cite{korringa50}.
Since thermally excited particle-hole pairs cause the Korringa
relaxation, $\gamma$ is proportional to the temperature $T$.
One can easily show that this term causes energy
relaxation due to transition between singlet and triplet states and
dephasing due to transition among three degenerate triplet states.

The last term of r.h.s. in Eq.\eqref{eq:MasterEq},
which is intrinsic to the system with the RKKY interaction,
is defined as
\begin{equation}
\begin{aligned}
\mathcal{D}_{\mathrm{ex}}\rho(t)
 &=
 -\dfrac{2\gamma_{\mathrm{ex}}}{\hbar}
 \Bigl(\left\{\vec{S}_1\cdot\vec{S}_2,\rho(t)\right\}
 -\dfrac{1}{2}\bigl(S_1^+\rho(t)S_2^-
\\
 &
\!
 +
\! S_1^-\rho(t)S_2^+
\!
 +
\! 2 S_1^z\rho(t)S_2^z
 +(\text{spin }1\leftrightarrow 2)\bigr)
 \Bigr).
 \label{eq:super-op-Dex}
\end{aligned} 
\end{equation}
This term exists only when the two localized spins 
interact with each other via conduction electrons.
The coefficient
$\gamma_{\mathrm{ex}}$ is expressed as $\gamma_{\mathrm{ex}}=\alpha_d k_B T
G_d(2 k_F R)$.
The Range function $G_{d}(2k_FR)$ is given in table \ref{tb:kappa-gamma},
and plotted by solid lines in Figs.\ref{fig:rangefunc}(a),(b), and (c).
As we shall show later, this term suppresses the energy relaxation
caused by singlet-triplet transitions due to the term $\mathcal{D}_{p}$.
The origin of this term is interference among
thermally exited particle-hole pairs near the Fermi surface.
Therefore, the range function $G_{d}(2k_FR)$ oscillates with the same
period as $F_{d}(2k_FR)$.
Reflecting the difference in physical mechanisms
behind them, the amplitude of $G_d(2k_FR)$ decays as $1/R^{d-1}$ while
that of $F_d(2k_FR)$ decays as slowly as $1/R^{d}$.
Especially for the one-dimensional electron gas,
the amplitude of the oscillation of $G_d(2k_FR)$ 
does not decay as $R$ increases.






Next, in order to capture the physical meaning of the term
$\mathcal{D}_{\mathrm{ex}}\rho(t)$, we examine the time evolution of
the expectation value
$\ave{\vec{S}_1\cdot\vec{S}_2}=\tr\{\rho(t)\vec{S}_1\cdot\vec{S}_2\}$, 
which is proportional to the energy expectation value of the RKKY interaction.
By using Eqs.\eqref{eq:MasterEq}-\eqref{eq:super-op-Dp},
we can show that it decays exponentially as
$\ave{\vec{S}_1\cdot\vec{S}_2}\propto\exp(-4(\gamma-\gamma_{\mathrm{ex}})t/\hbar)$.
Then the characteristic time of energy relaxation
of the RKKY interaction is given by
$\tau_{\mathrm{RKKY}}=\hbar/4(\gamma-\gamma_{\mathrm{ex}})$.
Since the term $\mathcal{D}_{\mathrm{ex}}\rho(t)$ is proportional to
$\gamma_{\mathrm{ex}}$, 
we can say that
the term $\mathcal{D}_{\mathrm{ex}}\rho(t)$ 
tends to suppress energy relaxation.
Note that $\gamma_{\mathrm{ex}}$ does not exceeds $\gamma$.
Especially when $\gamma_{\mathrm{ex}}=\gamma$,
the singlet-triplet transitions due to the term $\mathcal{D}_p$
does not occur and energy relaxation of the RKKY interaction
is completely suppressed.
However, 
the transitions among the triplet states are not suppressed
and dephasing rate remains finite
even if $\gamma_{\mathrm{ex}}=\gamma$.
As shown in Figs.\ref{fig:rangefunc}(a), (b), and (c),
the range function $G_d(2k_FR)$ takes its maximal values
when the $F_d(2k_FR)$ is negative.
Consequently the energy relaxation of the RKKY interaction
is suppressed when the RKKY interaction forms ferromagnetic coupling 
($J_\mathrm{RKKY}<0$).

\begin{figure}[]
 \begin{center}
  \includegraphics[width=0.85\linewidth,clip]{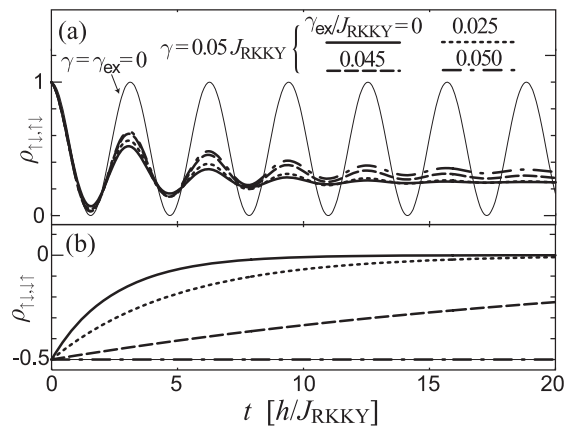}
 \end{center}
 \caption{(a) The time evolution of
 $\rho_{\uparrow\downarrow,\uparrow\downarrow}(t)$.
 The initial state is taken to be $\ket{\uparrow\downarrow}$ and the
 time is normalized by $h/J_{\mathrm{RKKY}}$.
 (b) The same plot for
 $\rho_{\uparrow\downarrow,\downarrow\uparrow}(t)$ with the initial
 state  $(\ket{\uparrow\downarrow}-\ket{\downarrow\uparrow})/\sqrt{2}$. For
 both panels, thin solid line represents the result for
 $\gamma=\gamma_{\mathrm{ex}}=0$ while all the other lines are
 for $\gamma=0.05J_{\mathrm{RKKY}}$.
 The parameter $\gamma_{\mathrm{ex}}$ is 
 taken to be 0, 0.025, 0.045,
 and 0.05$J_{\mathrm{RKKY}}$ for the thick solid, dotted, dashed, dot-dashed
 lines, respectively.
 }\label{fig:plot_dynamics}
\end{figure}

Figure \ref{fig:plot_dynamics}(a) shows the time evolution of the
diagonal element $\rho_{\uparrow\downarrow,\uparrow\downarrow}(t)$ of
the reduced density matrix.  The initial state is taken to be
$\ket{\uparrow\downarrow}$ and the time scale is normalized by
$h/J_{\mathrm{RKKY}}$.
The diagonal element $\rho_{\uparrow\downarrow,\uparrow\downarrow}(t)$ represents
the occupation probability of the state $\ket{\uparrow\downarrow}$.
Without decoherence, $\gamma=\gamma_{\mathrm{ex}}=0$,
the quantum state of the two-spin system oscillates between
$\ket{\uparrow\downarrow}$ and $\ket{\downarrow\uparrow}$
coherently and $\rho_{\uparrow\downarrow,\uparrow\downarrow}(t)$ 
shows a clear oscillation like a trigonometric function as indicated
by the thin solid line in Fig.\ref{fig:plot_dynamics}(a). 
In the case of $\gamma=0.05J_{\mathrm{RKKY}}$ and
$\gamma_{\mathrm{ex}}=0$, where each spins decohere independently, 
$\rho_{\uparrow\downarrow,\uparrow\downarrow}(t)$
shows damped oscillation due to $\mathcal{D}_p$ (thick solid line).
The dotted, dashed, and dot-dashed lines in
Fig. \ref{fig:plot_dynamics}(a) show the results for
$\gamma_{\mathrm{ex}}$=0.025, 0.045, and 0.05$J_{\mathrm{RKKY}}$, respectively.
The value of $\gamma_{\mathrm{ex}}$ can be controlled
by changing the distance between localized
spins, $R$.
One can see that the oscillation amplitude of
$\rho_{\uparrow\downarrow,\uparrow\downarrow}(t)$ decays more slowly
as $\gamma_{\mathrm{ex}}$ increases since the energy relaxation of
the RKKY interaction is
suppressed by $\gamma_{\mathrm{ex}}$.
Although energy relaxation of the RKKY interaction
is completely suppressed when $\gamma_{\mathrm{ex}}=\gamma$,
dephasing due to the transition among the triplet states
occurs as indicated by the dot-dashed line in
Fig. \ref{fig:plot_dynamics}(a).

In Fig. \ref{fig:plot_dynamics}(b) we plot the off-diagonal element 
$\rho_{\uparrow\downarrow,\downarrow\uparrow}(t)$
with the initial state
$(\ket{\uparrow\downarrow}-\ket{\downarrow\uparrow})/\sqrt{2}$ 
(the singlet state, i.e., the maximally entangled state of the
localized spins).
Because the singlet state is one of the energy eigenstates
of the RKKY interaction,
$\rho_{\uparrow\downarrow,\downarrow\uparrow}(t)$
is the conserved quantity when $\gamma=\gamma_{\mathrm{ex}}=0$.
If $\gamma$ takes a finite value, the singlet state changes to the
mixed state and $\rho_{\uparrow\downarrow,\downarrow\uparrow}(t)$ decays
exponentially to zero.
As shown in Fig.\ref{fig:plot_dynamics} (b), 
$\rho_{\uparrow\downarrow,\downarrow\uparrow}(t)$ decays 
more slowly as $\gamma_{\mathrm{ex}}$ increases, which means that
$\gamma_{\mathrm{ex}}$ suppresses decoherence of the singlet state. 
Especially in the case of $\gamma=\gamma_{\mathrm{ex}}$, the singlet state
remains as it is forever and the off-diagonal element
$\rho_{\uparrow\downarrow,\downarrow\uparrow}(t)$ never decays although
the localized spins always interact with the Fermion bath of
conduction electrons.
One can easily show that for the singlet state
the contributions of
$\mathcal{D}_{\mathrm{ex}}\rho(t)$ and $\mathcal{D}_p\rho(t)$ in
Eq.\eqref{eq:MasterEq} cancel out  when $\gamma=\gamma_{\mathrm{ex}}$.

Finally, we would like to discuss
the physical realization of a quantum gate using the RKKY interaction
and estimate the decoherence time.
Let us consider the system consisting of 
two quantum dots embedded in a 2DEG of
GaAs/AlGaAs heterostructure. We assume that the electron density of
the 2DEG is $n^{2\mathrm{D}}=7.3\times10^{11}\mathrm{cm}^{-2}$,
the Fermi energy and the Fermi wavelength are
$E_F\sim26 \mathrm{meV}$ and  $2\pi/k_F\sim 29.4 \mathrm{nm}$.
We also assume that
the charging energy of the quantum dot is $U=1.9\mathrm{meV}$,
the Lorentzian broadening of the localized-state
energy with full width at half maximum (FWHM) is
$\Gamma=295\mu\mathrm{eV}$,
and the temperature is $T=100\mathrm{mK}$
\cite{goldhaber-gordon98,kogan03}.
Each quantum dot contains a single localized spin which acts as a qubit.
The coupling between localized spins, i.e., the quantum gate operation is
controlled by applying a gate voltage to 2DEG.
For example the $\sqrt{\mathrm{SWAP}}$ gate operation\cite{loss98,nielsen},
which is known as a universal 2-bit quantum gate operation,
can be carried out if strength
of the RKKY interaction $J_{\mathrm{RKKY}}$ is controlled such that
$\int dt J_{\mathrm{RKKY}}(t)/\hbar=\pi/2$.
We suppose that the
inter dot distance is $R=\tilde{z}_2/k_F\sim 11\mathrm{nm}$ 
($\tilde{z}_2=2.40$ is the first zero-point of the Bessel function $J_0$),
that is, $\gamma_{\mathrm{ex}}=0$ and 
the energy relaxation is not suppressed.
The strength of the RKKY interaction takes $J_{\mathrm{RKKY}}\sim
3.4\mathrm{\mu eV}$ and the operation time defined as
$\tau_{\mathrm{op}}\equiv\pi\hbar/2J_{\mathrm{RKKY}}$
is $\sim 0.3\mathrm{ns}$.
Since $\gamma_{\mathrm{ex}}$ is assumed to be zero,
the decoherence time is determined only by $\gamma$
and is estimated as 
$\tau_{\mathrm{dec}}\equiv\hbar/\gamma\sim 62\mathrm{ns}$.
Therefore, about $200$ times coherent $\sqrt{\mathrm{SWAP}}$ operation
can be achieved.

The decoherence of a localized spin due to phonon scattering is suppressed in
a small quantum dot and long coherence time (order of millisecond) of
a single spin was observed in self-assembled semiconductor quantum
dots\cite{Kroutvar04}.  Therefore, the decoherence due to the Fermion
bath of conduction electrons is dominant in this system.  

More coherent quantum gate operation is available if we locate the
quantum dots very close to each other($k_FR\ll 1$).
In this case the value of $\gamma_{\mathrm{ex}}$ approaches $\gamma$ and
therefore we can take an advantage of the 
suppression of the energy relaxation.
Furthermore we can also obtain very strong 
ferromagnetic RKKY interaction\cite{tamura04}.
Experimentally quantum dot arrays with
$3$-$10\mathrm{nm}$ diameter quantum dots
have already been fabricated\cite{ueno01,shklyaev02a,shklyaev02b}.
We believe that the system we consider can be 
realized by using such quantum dot arrays
with a few $\mathrm{nm}$ inter dot distance.


In conclusion,
we have derived the kinetic equation for the system of two localized spins
embedded in an electron gas and 
have shown that particle-hole excitations in the
electron gas cause not only the RKKY interaction
but also decoherence of the two-spin system.
We also show that the strength of decoherence as well as
the RKKY interaction
strongly depends on the distance between two spins,
and energy relaxation due to singlet-triplet transition is suppressed
when the RKKY interaction is ferromagnetic ($J_\mathrm{RKKY}<0$).
We also estimate the decoherence time of the system
consisting of two quantum dots embedded in a two dimensional electron
gas(2DEG) to be $\tau_{\mathrm{dec}}\sim 60\mathrm{ns}$ within which
$\sim 200$ coherent $\sqrt{\mathrm{SWAP}}$ operations can be achieved.

We acknowledge H. Ebisawa for critically reading the manuscript.  This work was supported by CREST, MEXT.KAKENHI(No. 14076204 and No. 16710061), 
NAREGI Nanoscience Project, and NEDO Grant.


\end{document}